\documentclass[a4paper,onecolumn,11pt]{quantumarticle}
\pdfoutput=1
\usepackage[utf8]{inputenc}
\usepackage[english]{babel}
\usepackage[T1]{fontenc}
\usepackage{amsmath}
\usepackage{tabularx}
\usepackage{lipsum}
\usepackage{braket}
\usepackage[colorlinks=true, allcolors=blue]{hyperref}

\usepackage{amsmath, amssymb}
\usepackage{graphicx}
\usepackage{array}         
\usepackage{booktabs}      
\usepackage{caption}
\usepackage{hyperref}

\newcommand{\beit}{\affiliation{BEIT sp. z o.o., Mogilska 43, 31-545 Kraków, Poland}}%
\newcommand{\beitemail}{\email[]{emil@beit.tech}}%
\newcommand{\beitemailsp}{\email[]{szymonzdzislawplis@gmail.com}}%
\newcommand{\beitweb}{\homepage{https://www.beit.tech}}%

\begin{document}

\title{Quantum Discrete Variable Representations}

\author{Szymon Pliś}\beit\beitemailsp
\orcid{0000-0002-2445-2701}
\author{Emil Zak}\beitemail\beitweb\beit
\orcid{0000-0003-3528-4907}

\maketitle

\begin{abstract}
We present a fault-tolerant quantum algorithm for implementing the Discrete Variable Representation (DVR) transformation, a technique widely used in simulations of quantum-mechanical Hamiltonians. DVR provides a diagonal representation of local operators and enables sparse Hamiltonian structures, making it a powerful alternative to the finite basis representation (FBR), particularly in high-dimensional problems. While DVR has been extensively used in classical simulations, its quantum implementation, particularly using Gaussian quadrature grids, remains underexplored. 
We develop a quantum circuit that efficiently transforms FBR into DVR by following a recursive construction based on quantum arithmetic operations, and we compare this approach with methods that directly load DVR matrix elements using quantum read-only memory (QROM). We analyze the quantum resources, including T-gate and qubit counts, required for implementing the DVR unitary and discuss preferable choices of QROM-based and recursive-based methods for a given matrix size and precision. 
This study lays the groundwork for utilizing DVR Hamiltonians in quantum algorithms such as quantum phase estimation with block encoding.
\end{abstract}

\section{Introduction}
The Discrete Variable Representation (DVR) technique \cite{Harris1965, Light1987, Light1989, Light2000} formulates partial differential equations using a localized basis associated with an underlying spatial grid. DVR is constructed as a linear combination of variational basis functions such that, in the spatial representation, each basis function is localized around the grid point it is associated with.

DVRs are widely used in computational physics, simulations, and engineering, with successful applications in nuclear motion theory \cite{DVR3D, Wang2009, Fbri2017}, scattering calculations \cite{Colbert1992}, nuclear physics \cite{Bulgac2013}, electronic structure theory \cite{Parrish2013}, attosecond physics \cite{Demekhin2013}, and quantum circuit simulations \cite{Richman}.

In this work, we focus on Gaussian discrete variable representations in the context of quantum simulations of Hamiltonians. A commonly used approach for solving the associated Schrödinger equation is the finite basis representation (FBR), in which the Hamiltonian operator is represented in a truncated variational basis, and matrix elements are evaluated via quadrature~\cite{Light2000}. The DVR can be obtained via a unitary transformation of the FBR and offers several advantages.

First, DVR yields diagonal representations for local quantum-mechanical operators such as multi-dimensional potential energy surfaces (PES)~\cite{Light2000}. For this reason, DVR is often preferred over variational bases, particularly when evaluating many-body potential energy matrix elements is computationally intensive. Typically, the kinetic energy operator, being a second-order differential operator, has a simpler structure than the potential. When expressed in a DVR basis, the potential energy becomes a diagonal matrix with entries corresponding to the potential evaluated at DVR grid points. The kinetic energy operator is first evaluated analytically in a variational basis and then transformed to the DVR basis.
Such an approach often results in sparse Hamiltonian matrices, making DVR especially valuable in molecular physics, where the curse of dimensionality imposes severe computational limits. Another notable advantage of DVR is the ability to selectively prune basis functions based on physical insight, thanks to their spatial localization. Finally, in contrast to finite-difference grid-based methods, DVRs exhibit exponential convergence in accuracy of approximating bound states of many-body Hamiltonians~\cite{Littlejohn2002}.

For multi-dimensional problems, DVR can be extended using a direct-product basis \cite{Light2000}. On classical computers, the memory required for this transformation scales exponentially with the number of dimensions, as $N^D$, where $N$ is the number of basis functions per dimension and $D$ is the dimensionality. Quantum computers however can reduce this memory scaling to $\mathcal{O}(ND)$. 
Beyond memory advantages, DVR may also offer quantum-computational speedups due to the sparsity of the DVR Hamiltonian~\cite{camps}. We therefore argue that, given an efficient circuit implementation of the FBR-to-DVR transformation, Hamiltonian simulation problems can be solved  on quantum computers at a lower gate cost in the DVR basis than in the FBR basis.

A particularly appealing application of DVR is in computing Hamiltonian eigenvalues. In this context, fault-tolerant quantum algorithms such as Quantum Phase Estimation (QPE) often rely on block encoding of the Hamiltonian~\cite{babbush2018, lee2021, low2019, gilyen2019, Guzik2024}. The norm of the Hamiltonian, which proportionally scales the T-gate complexity of QPE, can be reduced compared to variational methods through appropriate truncations, such as removing irrelevant DVR grid points. Moreover, the Hamiltonian can often be decomposed into a diagonal PES and a structured kinetic energy operator, where the diagonal part can be efficiently block-encoded using standard techniques~\cite{camps,stick,gosset}.

Several quantum algorithms have been developed for grid-based physical simulations \cite{Macridin2018PRL, Macridin2018PRA, Ollitrault2023, Chan2023} and for solving differential equations \cite{Childs2021, TostiBalducci2022}, with applications in electronic structure \cite{Su2021, Chan2023} and nuclear motion calculations \cite{Guzik2024, Ollitrault2023, Lee2022}. A quantum algorithm for wavelet-based representations was also proposed in Ref.~\cite{Bagherimehrab2024}. For Hamiltonian simulation in general, the widely used quantum Fourier-transformed representation based on equidistant grids provides several advantages~\cite{Su2021, Guzik2024}. However, quantum simulations using non-equidistant grids, particularly those defined by Gaussian quadrature, which allows exact integration of polynomials up to a given degree, have not yet been explored in detail. In particular, a quantum circuit implementing the FBR-to-DVR unitary transformation could enable more efficient simulations in DVR.

Below we present a quantum algorithm for implementing the FBR-to-DVR transformation and for loading DVR matrix elements into qubit states. In Section~\ref{sec:gaussian-dvr}, we review the fundamentals of Gaussian DVRs, which are used throughout the paper. In Section~\ref{sec:DVRoracle-rec}, we describe the construction of a quantum DVR oracle used to implement the FBR-to-DVR transformation, which is detailed in Section~\ref{sec:DVRunitary}. For each construction, we estimate fault-tolerant quantum resources, including T-gate counts, circuit depth, and qubit requirements. Finally, we briefly discuss potential applications of DVRs on quantum devices.

\section{Gaussian DVRs}
\label{sec:gaussian-dvr}
Gaussian DVR matrix elements can be defined as:
 \begin{equation}
      T_{pq}=N_q\sqrt{w_p}p_q(x_p), \hbox{ for } p,q=0,\ldots,N-1 
      \label{eq:dvr-definition}
 \end{equation}
where $N_q=\|p_q\|_{L^2(\mu)}^{-1}$ is the normalization constant, $w_p$ are the Gaussian quadrature weights, $p_q(x_p)$ is the value of the degree-$q$ orthogonal polynomial defining the Gaussian quadrature evaluated at the Gaussian quadrature node $x_p$~\cite{Stoer1980}. Gaussian quadrature nodes and weights are chosen such that the integral:
\begin{equation}
    \int fd\mu \approx \sum_{k=0}^{N-1} w_kf(x_k)
\end{equation}
is exact for $f$ chosen as polynomials of degree less or equal $2N-1$.
The Gaussian quadrature nodes $x_0$, $x_1$, $\ldots$, $x_{N-1}$ are chosen as zeroes of $p_N$. The weights are then determined from the equation: $$w_k=\frac{a_N}{a_{N-1}}\frac{\|p_{N-1}\|_{L^2(\mu)}}{p_N'(x_k)p_{N-1}(x_k)}\geq0, \;k=0,\ldots,N-1,$$

Orthogonal polynomials associated with Gaussian quadratures satisfy the three-term recurrence relations:
\begin{equation}
p_{q}(x)=(a_q+b_qx)p_{q-1}(x)+c_qp_{q-2}(x)
    \label{eq:three-term}
\end{equation} 
which gives the following recurrence for the columns of the DVR matrix
\begin{equation}
T_{pq}=(A_q+B_qx_p)T_{pq-1}+C_qT_{pq-2}.
    \label{eq:dvr-recurrence}
\end{equation}
In eq.~\ref{eq:dvr-recurrence}, $A_q = \frac{N_q}{N_{q-1}}a_q$, $B_q = \frac{N_q}{N_{q-1}}b_q$ and $C_q = \frac{N_q}{N_{q-2}}c_q$.

For simulations of quantum-mechanical Hamiltonians states can be represented using Finite Basis Representation (FBR), given by a linear combination of basis functions: 
\begin{equation}
\psi(x) = \sum_{j=0}^{N-1}c_j\phi_j(x)
    \label{eq:fbr}
\end{equation}
where the coefficients $c_j$ are determined variationally.
The DVR basis functions $d_p(x)$ can be written as a linear combination of FBR basis functions:
 \begin{equation}
d_p(x)=\sum_{q=0}^{N-1}T_{pq}\phi_q(x)
\end{equation}
 Note that DVR functions vanish in all quadrature grid points except one, which can be shown by evaluating the DVR wavefunction at $k$-th grid point:
 \begin{equation}
d_l(x_k)=\sum_{j=0}^{N-1}\sqrt{w_l\omega(x_k)}p_j(x_l)p_j(x_k).
\end{equation}
From orthogonality of $p_j(x)$ we find
 \begin{equation}
d_l(x_k)=\sqrt{\frac{\omega(x_k)}{w_l}}\delta_{lk}
\end{equation}

The DVR-to-FBR transformation $\mathbf{T}$ is unitary because the Gaussian quadrature approximation to the overlap integral: $\int p_i(x)p_j(x)dx \approx (\mathbf{T}^{\dag}\mathbf{T})_{ij}=\sum_{k=1}^N w_k p_i(x_k)p_j(x_k)$ is exact for $i$ and $j$ both smaller or equal $N-1$, i.e. their combined degree is maximally $2N-2$. 
For this reason $\mathbf{T}^{\dag}\mathbf{T} = \mathbf{T}\mathbf{T}^{\dag}=\mathbf{1}$.
Note that the left index $p$ in the DVR matrix given in eq.\ref{eq:dvr-definition} refers to grid-point (physical space) while the right index $q$ refers to variational basis function index. Thus the DVR transformation connects delocalized (global) basis functions space with localized functions associated with a grid.

For solving quantum-mechanical problems, DVR is often constructed based on orthogonal polynomials that solve the Schrödinger equation for specific model systems: Hermite polynomials for the harmonic oscillator, Legendre polynomials for spherically symmetric problems, Laguerre polynomials for the hydrogen atom, associated Laguerre polynomials for the Morse oscillator, Chebyshev polynomials for a particle in a square potential well, and Lobatto polynomials for problems with fixed boundary conditions~\cite{abramowitz,Light2000}. An example set of DVR basis functions defined for the Hermite polynomias is shown in Figure~\ref{Fig:DVRbasis}. Further details about Gaussian DVRs and methods for generating them are discussed in Appendix~\ref{sec:AppendixA}.

\begin{figure}
\begin{center}
  \includegraphics[width=\linewidth]{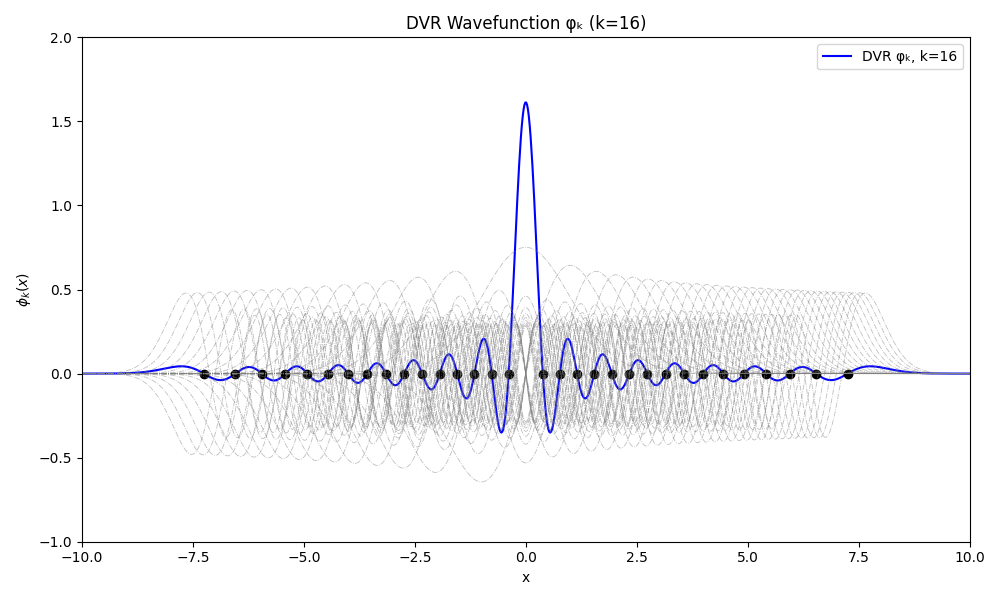}
  \caption{Example DVR basis function (blue solid line) constructed from 33 Harmonic oscillator wavefunctions (grey dashed lines). Zeros of the DVR basis function marking the Gauss-Hermite quadrature nodes are marked.}
  \label{Fig:DVRbasis}
  \end{center}
\end{figure}

\section{Quantum DVR Oracle}
\subsection{Constructing the DVR oracle with the recursive method}
\label{sec:DVRoracle-rec}
For implementing the FBR-DVR unitary circuit we first construct a unitary, which we call the DVR oracle. The DVR oracle is defined as quantum state transformation encoding the elements of the DVR matrix $T$ in qubit registers, written as:
\begin{equation}
    \mathcal{T}\ket p \ket q \ket 0=\ket p \ket q \ket{ T_{pq}}
    \label{eq:DVRoracle-eq}
\end{equation}
where $T_{pq}$ are the DVR matrix elements, $\ket{q}$ is argument state representing column index (basis state index) of the DVR matrix, $\ket{p}$ is argument state representing row index (grid point index).
We represent the column index as $q=wF+v$, where $w=0,1,2,...,\frac{N}{F}-1$ and $v=0,1,2,...,F-1$ for the purpose of splitting the column space in the DVR matrix into $\frac{N}{F}$ segments. Let us call $F$ the \textit{segmentation parameter}. The overall construction is divided into several steps, described in the following paragraphs.

\paragraph{Segment initialization.}
The construction begins with initialization unitary that loads $2\frac{N}{F}$ columns of the DVR matrix using quantum random-access memory (QROM) written as:
\begin{equation}
    \hat{U}^{(init)}|p\rangle | w\rangle \ket{0} |z\rangle |z\rangle =|p\rangle | w\rangle\ket{x_p} |z\oplus T_{p\tilde{q}-1}\rangle |z\oplus T_{p\tilde{q}}\rangle
    \label{eq:uinit}
\end{equation}
where $\tilde{q}=wF+\frac{F}{2}$ is the midpoint index in each of the $\frac{N}{F}$ segments dividing the DVR matrix. State $\ket{x_p}$ keeps nodes of the $N$-th orthogonal polynomial defining the DVR.
This step requires loading $\frac{N^2}{F}$ pairs of numbers $(T_{p\tilde{q}-1},T_{p\tilde{q}})$, given by $m$-bits each, leading to the T-gate cost of 
\begin{equation}
    C(\hat{U}_{init})=2\frac{N\sqrt{m}}{\sqrt{F}}+\sqrt{Nm}
    \label{eq:init_SELSWAP}
\end{equation}
in case of using the SELSWAP algorithm~\cite{Low2024} and 
\begin{equation}
    C(\hat{U}_{init})=\frac{N^2}{F} +N
    \label{eq:init_SEL}
\end{equation}
for the SELECT QROM.

\begin{figure}[h!]
    \centering
    \includegraphics[width=0.3\linewidth]{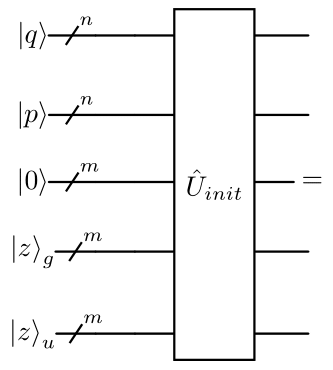}\\
    \includegraphics[width=0.8\linewidth]{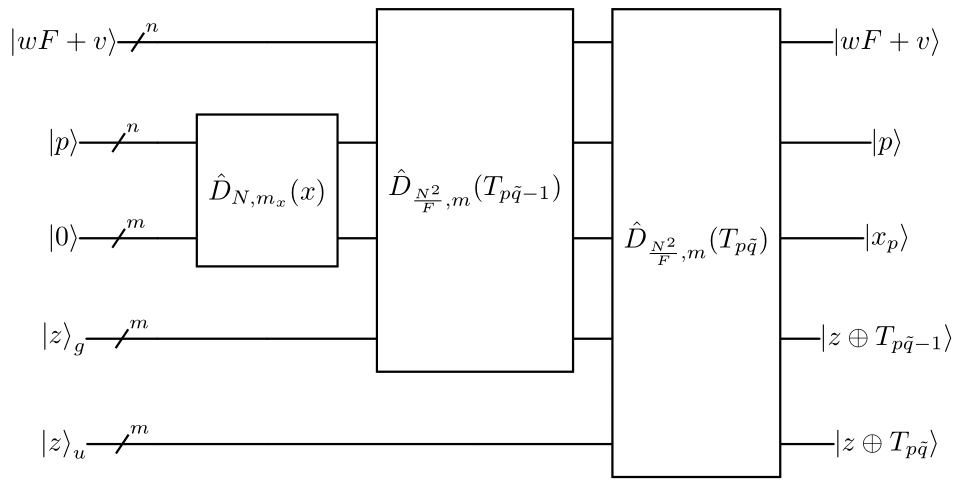}
    \caption{Quantum circuit representing initialization unitary for the DVR oracle defined in eq.~\ref{eq:uinit}. $\hat{D}_{N,m}(A)$ denotes QROM encoding data represented by function $A$ for $N$ arguments and output stored in $m$ qubits. Here $T_{p\tilde{q}}$ denotes the DVR matrix element for the $\tilde{q}$'th column, where $\tilde{q}=wF+\frac{F}{2}$. The column index $q$ is represented as $q=wF+v$.}
    \label{fig:DVR_initialization}
\end{figure}

\begin{figure}[h!]
    \includegraphics[width=1\linewidth]{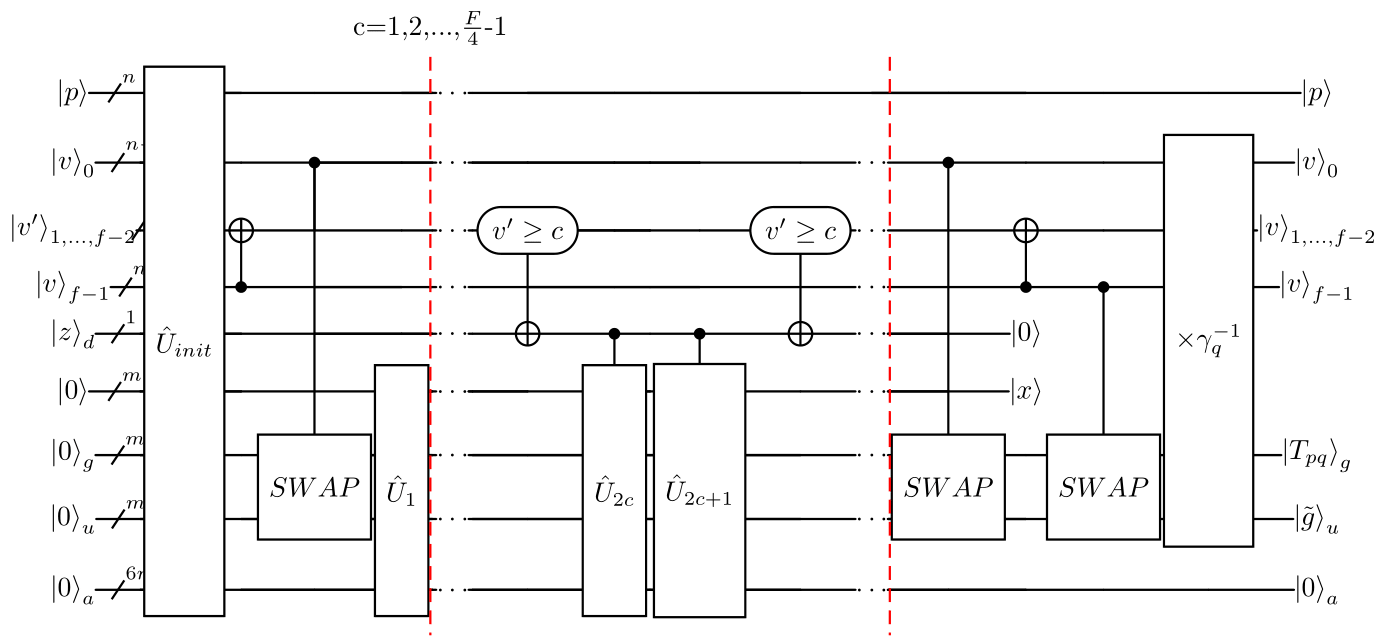}
    \caption{Quantum circuit representing the DVR oracle defined in eq.~\ref{eq:DVRoracle-eq}. $\hat{U}_{init}$ represents state initialization oracle shown in Fig.~\ref{fig:DVR_initialization}.  $\hat{U}_c$ are iteration unitaries shown in Fig.~\ref{fig:arithmetic}. $\gamma_q^{-1}$ is a gate multiplying the result by the appropriate scaling factor defined in eq.~\ref{eq:gamma}. The result is returned in $m$-qubit register $\ket{T_{pq}}_g$.  } 
    \label{fig:DVRoracle}
\end{figure}

\paragraph{Recursive construction.}
Following the initialization step, the algorithm performs a sequence of operations $\hat{U}_{2c}$, $\hat{U}_{2c+1}$ controlled by the running index values $c=1,2,...,\frac{F}{4}-1$ enumerating the range for the segmentation parameter $F$, as depicted in Figure~\ref{fig:DVRoracle}. 
The recursive block of the circuit is wrapped with SWAP operations applied to two output registers, $m$-qubit  each. The SWAP operations are controlled on the most significant bit $v_0$ of $v = v_02^{f-1}+v_12^{f-2}+...+v_{f-1}2^0$, where $v=0,1,2,...,F-1$ and the column index is given by $q=wF+v$. Here $F=2^f$.  The output registers keep the intermediate values in the recursive construction procedure, whereas the role of the SWAP operations is to move the output to the $\ket{.}_g$ register, regardless of the parity of the column index queried. The cost of the single controlled-SWAP operation is $m$ Toffoli gates. 

\begin{figure}[h!]
    \centering
     \includegraphics[width=0.6\linewidth]{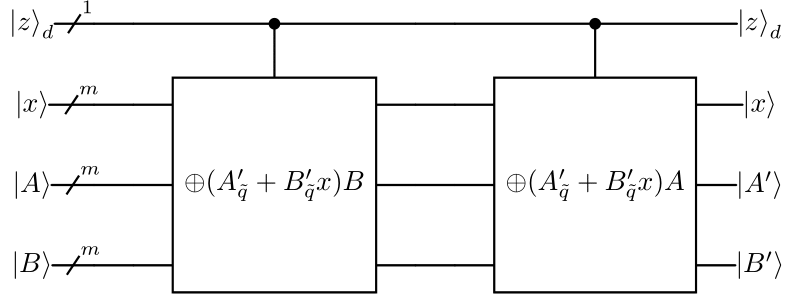}
        \caption{Quantum circuit representing  $\hat{U}_c$ iteration unitary in the the DVR oracle. $\ket{z}_d$ is the \textit{dump} qubit register shown also in Figure~\ref{fig:DVRoracle}.}
    \label{fig:arithmetic}
\end{figure}

\newpage
\begin{figure}[h!]
\includegraphics[width=1\linewidth]{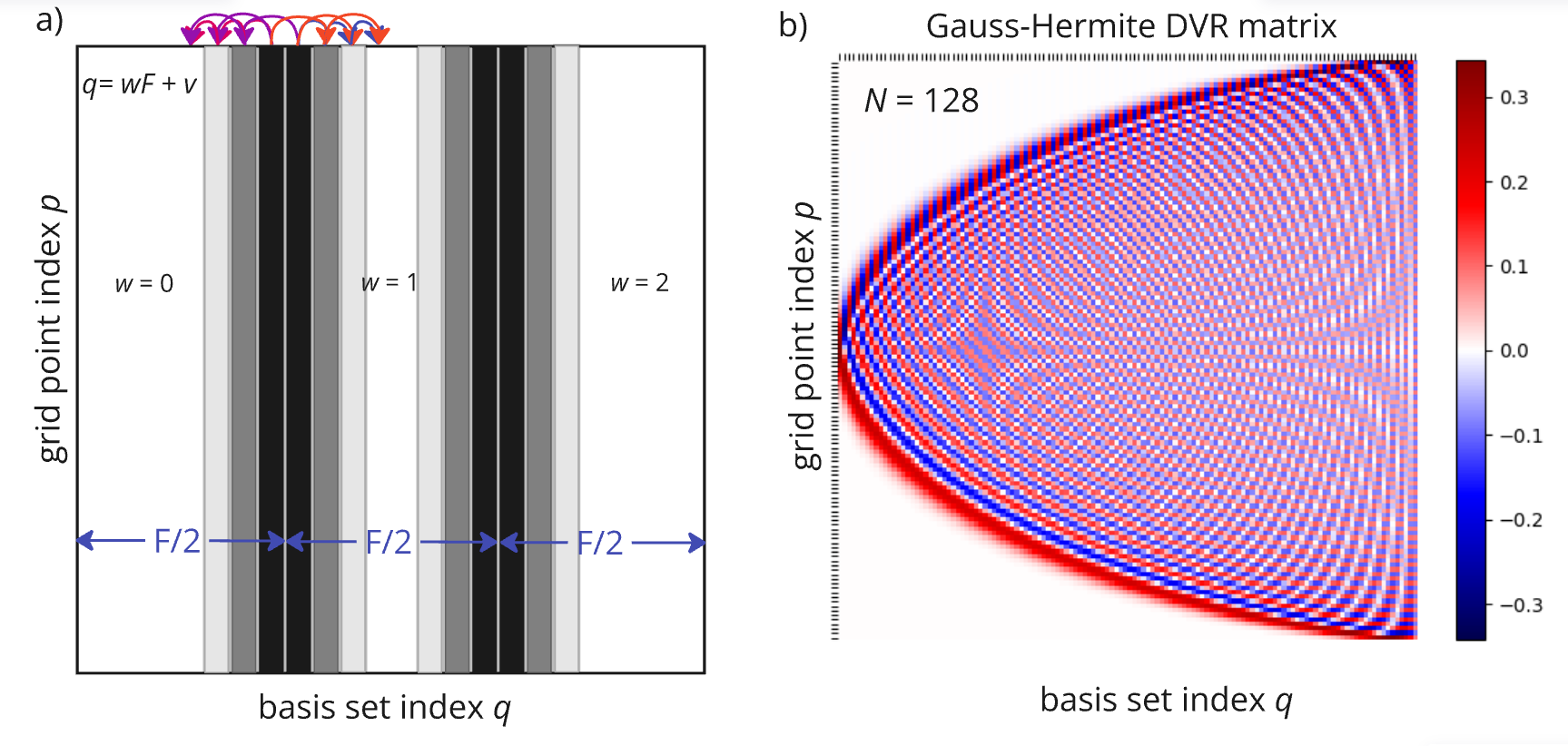}
    \caption{a) Schematic representation for the segment construction of the DVR oracle matrix. Basis set (column) index is represented as $q=wF+v$, where $w=0,1,2,...,\frac{N}{F}-1$ and $v=0,1,2,...,F-1$ for the purpose of splitting the column space in the DVR matrix into $\frac{N}{F}$ segments. Columns in each segment are constructed simultaneously in the ascending and descending horizontal direction following the three-step recursion given in eq.~\ref{eq:recursion}; b) Color map encoding values of the DVR transformation matrix for the Gauss-Hermite quadrature with $N=128$ elements.}
    \label{fig:schematic}
\end{figure}

The recursive block is completed upon $\frac{F}{4}-1$ steps (marked between red vertical dashed lines in Figure~\ref{fig:DVRoracle}), where in each step two columns per segment are constructed. The columns are successively built moving in the descending and ascending order of $q$ values simultaneously, starting from the midpoint columns indexed by $\tilde{q}=wF+\frac{F}{2}$. Schematic for this procedure is shown in Figure~\ref{fig:schematic}a. In constructing $\hat{U}_{2c}$ and $\hat{U}_{2c+1}$ we utilize the three-term recurrence relation satisfied by the DVR matrix elements:
\begin{equation}
    T_{p\;q+2}=(A_q+B_qx_p)T_{p\;q+1}+C_qT_{pq}.
    \label{eq:recursion}
\end{equation}
where the specific values of coefficients $A_q,B_q,C_q$ depend on the underlying orthogonal polynomial. 
The number of effective parameters in eq.~\ref{eq:recursion} can be reduced by introducing scaled matrix elements: $T'_{p\;q+2}=\gamma_q T_{p\;q+2}$, giving the following recursions for the ascending and descending directions, respectively:
\begin{align}
    T'_{pq}=(A'_q+B'_qx_p)T'_{p\;q-1}+T'_{p\;q-2} \qquad \hbox{for} \quad q=wF+\frac{F}{2}+k \\
    T'_{pq}=(A'_q+B'_qx_p)T'_{p\;q+1}+T'_{p\;q+2} \qquad \hbox{for} \quad q=wF+\frac{F}{2}-1-k
    \label{eq:recursion'}
\end{align}
where $k=1,\ldots,\frac{F}{2}-1$. For the two initial columns in each segment we set $\gamma_q=1$ (for $q=wF+\frac{F}{2}-1$ and $q=wF+\frac{F}{2}$).
For the ascending direction of $q=wF+\frac{F}{2}+k$ the scaling constant is given by 
\begin{equation}
\gamma_q=\gamma_{q-2}C_{q-2}^{-1}=(C_{q-2}C_{q-4}\ldots C_i)^{-1}
\label{eq:gamma}
\end{equation}
where $i\in\{wF+\frac{F}{2}-1,wF+\frac{F}{2}\}$ has the same parity as $q$. 
Then $(A'_q,B'_q)=\frac{\gamma_{q}}{\gamma_{q-1}}(A_{q-2},B_{q-2})$. For the descending direction  of $q=wF+\frac{F}{2}-1-k$, the scaling constant is given by the formula $\gamma_q=\gamma_{q+2}C_q=C_{q}C_{q+2}\ldots C_i$,  where again $i\in\{wF+\frac{F}{2}-1,wF+\frac{F}{2}\}$ has the same parity as $q$. In this case $(A'_q,B'_q)=-\frac{\gamma_{q+2}}{\gamma_{q+1}}(A_{q},B_{q})$. Each unitary iteration $\hat{U}_{2c}$ and $\hat{U}_{2c+1}$ involves two arithmetic blocks, two multiplications and one addition each, as shown in Figure~\ref{fig:arithmetic}. The appropriate qubit state transformation for $\hat{U}_{2c}$ and $\hat{U}_{2c+1}$ can be written as:
\begin{align}
U_{2c} \ket q\ket x\ket{A}\ket{B} &=
    \ket q\ket x\ket{A} \ket{B\oplus (A'_{\hat q}+B'_{\hat q}x)A}  
    \label{eq:uc-unitaries1}
    \\
U_{2c+1} \ket q\ket x\ket{A}\ket{B} &=    \ket q\ket x\ket{A\oplus (A'_{\check q}+B'_{\check q}x)B}\ket{B}    
\label{eq:uc-unitaries2}
\end{align}
for  $|v-\frac{F-1}{2}|>2c$. The indices for recursion constants appearing in eqs.~\ref{eq:uc-unitaries1},~\ref{eq:uc-unitaries2} are given by
\begin{equation}
    \hat q=\hat q(v_0,w,c)=\begin{cases}    
wF+F/2+2c \hbox{ for } v_0=1 
\\ 
wF+F/2-1-2c \hbox{ for } v_0=0
\end{cases}  
\end{equation}
 and
\begin{equation}
    \check q=\check q(v_0,w,c)=\begin{cases}    
wF+F/2+2c+1 \hbox{ for } v_0=1 
\\ 
wF+F/2-2-2c \hbox{ for } v_0=0.
\end{cases}  
\end{equation}
where $q=wF+v$. Further details of the quantum arithmetics are given in Appendix B.

\paragraph{Cost estimation for the recursive block.}
The recursive block given by eqs.~\ref{eq:uc-unitaries1},\ref{eq:uc-unitaries2} and shown between the red dashed lines in Figure~\ref{fig:DVRoracle} requires loading $2N/F$ constants $A'_q,B'_q$ with QROM, associated with the $2N/F$ Toffoli gate cost. The total Toffoli cost for the $U_c$ iteration shown in Figure~\ref{fig:arithmetic} consists of multiplication and addition costs (cf. eq.~\ref{eq:uc-unitaries1}\ref{eq:uc-unitaries2}). Circuit performing multiplication of numbers $B'_q$ and $x$, denoted $MUL(B'_q,x)$, involves $2m^2$ Toffoli gates. The result of multiplication is rounded to $m$-bits. Circuit performing addition of $A'_q$ and $B'_qx$, denoted as $ADD[A'_q,B'_qx]$, costs $4m$ Toffoli gates. All the above arithmetic operations must be uncomputed, which doubles the Toffoli cost. Finally, the bitwise addition in eq.~\ref{eq:uc-unitaries1}, $B\oplus (A'_{\hat q}+B'_{\hat q}x)A$ costs $2m$ Toffoli gates, giving the overall Toffoli count for the arithmetic part $\hat{U}_c$:
 \begin{align}
C(\hat{U}_c)=2C(\hat{D}_{\frac{N}{F},m}(A') ) +2C(\hat{D}_{\frac{N}{F},m}(B') ) + 2C(MUL(B'_q,x))+2C(ADD[A'_q,B'_qx]) + \\
+2C(MUL(A'_{\hat q}+B'_{\hat q}x,B))+C(B\oplus (A'_{\hat q}+B'_{\hat q}x)A) =  4\frac{N}{F} + 8m^2 + 10m
     \label{eq:cost_artihmetic}
 \end{align}
There are $\frac{F}{4}-1$ pairs of transformations ($\hat{U}_{2c},\hat{U}_{2c+1}$) giving the total cost of the recurisive block:
 \begin{equation}
C_{rec}=\left(\frac{F}{4}-1\right)\left(2C(\hat{U}_c)+2C(v'\geq c)\right)
     \label{eq:cost_recursive}
 \end{equation}
where the checking the condition $v'\geq c$ (on $f-2$ qubits) costs $f-2$ Toffoli gates, while SWAP operations cost $3m$ Toffolis in total. Here $v' = v_12^{f-2}+...+v_{f-2}2^1$ is a bitmask of $v$ with the most and the least significant bit removed. The cost of loading scaling constants $\gamma_q^{-1}$ and multiplication $C(MUL(A,\gamma_q^{-1}))$ shown in Figure~\ref{fig:DVRoracle} is $N+2m^2$ Toffoli gates, assuming SELECT QROM.

\paragraph{Resource estimation summary.}
We summarize the quantum computing resources for implementing the DVR oracle circuit given in eq.~\ref{eq:DVRoracle-eq}, by first estimating the required number of qubits. The DVR oracle requires $2n+m$ (nonancilla) qubits, $n$ per argument register keeping values of matrix row and column indices, respectively, plus $m$ qubits to represent the output matrix elements. We use additional $m$ qubits for the purpose of loading pairs of entries of the DVR matrix, as required by the recursive scheme adopted in eqs.~\ref{eq:uc-unitaries1},\ref{eq:uc-unitaries2}. Quantum arithmetic performed in eqs.~\ref{eq:uc-unitaries1},\ref{eq:uc-unitaries2} requires $6m$ additional ancilla qubits. Thus the total qubit count is $2n+9m$.

The total Toffoli gate cost for the DVR oracle consists of initalization cost $C(\hat{U}_{init})$, the cost of the recursive block $C_{rec}$ plus the cost of the remaining operations: $C(\hat{U}_{1})$, $C(SWAPS)$ and $C(MUL(A,\gamma_q^{-1}))$, giving:
 \begin{equation}
C_{ODVR}=C(\hat{U}_{init})+C(\hat{U}_{1})+C_{rec} +C(MUL(A,\gamma_q^{-1}))+C(SWAPS)  
     \label{eq:cost_recursive}
 \end{equation}
The total cost is thus:
 \begin{equation}
C_{ODVR}= \frac{N^2}{F} + N +   4\frac{N}{F}+8m^2+10m +\left(\frac{F}{4}-1\right)\left(16m^2+8m+4\frac{N}{F}+2(f-2)\right)+3m +N+2m^2
     \label{eq:DVRoracle-cost}
 \end{equation}
in which the dominant cost is given by
 \begin{equation}
C_{ODVR} \approx \frac{N^2}{F}+4Fm^2
     \label{eq:cost_recursive_gates_asymptotic}
 \end{equation}
In eqs.~\ref{eq:DVRoracle-cost}-\ref{eq:cost_recursive_gates_asymptotic} we assume SELECT QROM for initial data loading ($\hat{U}_{init}$), with the benefit of qubit saving compared to the SELSWAP approach. When one optimizes for balance between the number of qubits and T-gates, the SELSWAP version of QROM in $\hat{U}_{init}$ can be chosen (cf. eq.~\ref{eq:init_SELSWAP}) giving the asymptotic cost:
 \begin{equation}
C_{ODVR} \approx 
4Fm^2+N\left(\sqrt{\frac{m}{F}}+1\right)
\label{eq:cost_recursive_gates_asymptotic}
\end{equation}

{\bf The DVR oracle for parity-conserving orthogonal polynomials.} 
When $A_q = 0$ for all $q$, parity symmetry emerges in the orthogonal polynomials defined by the recursive relations shown in Eq.~\ref{eq:recursion}. For example, Hermite polynomials and Jacobi polynomials with $\alpha = \beta$ (such as Legendre and Chebyshev) exhibit the symmetry $T_{pq} = (-1)^q T_{N-p, q}$. These represent a special class of orthogonal polynomials that allow for additional quantum computing cost savings.

In such cases, it is sufficient to load only the upper half of the DVR matrix, applying a sign flip to the lower half for odd-numbered columns, at no additional Toffoli cost. Consequently, the Toffoli cost of QROM loading $\hat{D}_{N/2,m}(x)$ is halved in the SELECT method and reduced by a factor of $\sqrt{2}$ for equal-split SELSWAP. The cost of loading the initial $2\frac{N}{F}$ columns in $\hat{U}_{init}$ is reduced by the same factor. Additionally, the operations associated with adding the free term $A_{q'}$ in the recursion are removed, saving an additional $2m$ Toffoli gates per recursive step.

In summary, for parity-conserving polynomials, the total Toffoli gate cost of the DVR oracle is given by

\begin{equation} 
C_{ODVR}= \frac{N^2}{2F} + \frac{N}{2} +  4\frac{N}{F}+8m^2+10m+ \left(\frac{F}{4}-1\right)\left(16m^2 + 8m + 4\frac{N}{F} + 2(f - 2)\right) + 3m+N + 2m^2 \label{eq:DVRoracle-cost-sym} 
\end{equation}

\subsection{DVR Oracle via SELSWAP QROM} \label{sec:DVRoracle-QROM}
An alternative construction of the DVR oracle, compared to the \textit{recursive method} described in Section~\ref{sec:DVRoracle-rec}, employs the generic QROM technique. In this approach, matrix elements are loaded into the quantum computer using a combination of the SELECT and SWAP oracles introduced in Ref.~\cite{Low2024}. Loading the DVR matrix into a quantum state via the SELSWAP method requires $N\sqrt{m}$ Toffoli gates and $N\sqrt{m}$ ancilla qubits.

For parity-conserving orthogonal polynomials, this cost is reduced to $\frac{1}{2}N\sqrt{m}$ Toffoli gates, i.e. approximately half that of the general case. The $T$-depth of the SELSWAP method is roughly equal to the $T$-gate count, with the SELECT operation dominating the circuit depth. A comparison of the direct SELSWAP and recursive approaches is presented in the following section.

\subsection{Comparison of techniques for constructing the DVR oracle}
\label{sec:DVRoracle-compare}
The two approaches for constructing the DVR oracle discussed in sections ~\ref{sec:DVRoracle-rec} and~\ref{sec:DVRoracle-QROM} are summarized in Table~\ref{tab:comparison}.

\begin{table}[h]
    \centering
    \renewcommand{\arraystretch}{1.2}
    \setlength{\tabcolsep}{3pt} 
    \begin{tabularx}{\linewidth}{l >{\centering\arraybackslash}X >{\centering\arraybackslash}X >{\centering\arraybackslash}X >{\centering\arraybackslash}X}
        \toprule
        Quantity & T-count & T-depth & Qubit count & Volume \\
        \midrule
                \hline
        LKS & $N\sqrt{m}$ & $N\sqrt{m}$ & $N\sqrt{m}$ & $N^2m$ \\
        REC & $4Fm^2+\frac{N^2}{F}$ & $4Fm^2+\frac{N^2}{F}$ & $2n+9m$ & $36Fm^3+9m\frac{N^2}{F}$ \\
        REC-LKS & $4Fm^2++N\left(\sqrt{\frac{m}{F}}+1\right)$ & $4Fm^2++N\left(\sqrt{\frac{m}{F}}+1\right)$ & $N\sqrt{\frac{m}{F}}+\sqrt{Nm}++2n+6m$ & $N^2\frac{m}{F}+24Fm^3$ \\
        \bottomrule
    \end{tabularx}
    \caption{\scriptsize Comparison of methods for constructing the DVR oracle defined in eq.~\ref{eq:DVRoracle-eq}. LKS stands for the direct QROM method using the SELSWAP technique from ref.~\cite{Low2024}, REC is our present method where the initialization unitary given in eq.~\ref{eq:init_SEL} is constructed using SELECT QROM (saving ancilla qubits), while REC-LKS is the present method where the initial unitary given in eq.~\ref{eq:init_SELSWAP} is constructed using SELSWAP QROM (balancing T-gate cost with ancilla qubits). Only dominant terms are shown for the recursive method. The columns correspond to: T-count, T-depth, qubit count, and quantum volume (qubit count $\times$ T-count).}
    \label{tab:comparison}
\end{table}

\begin{figure}[h!]
\includegraphics[width=1\linewidth]{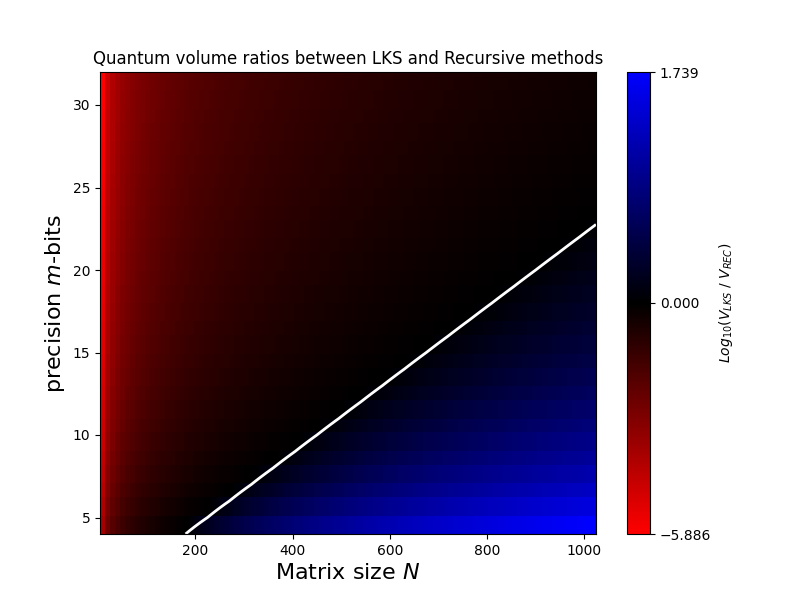}
    \caption{Comparison of quantum volumes for the LKS method and the recursive method for constructing the DVR oracle as a function of the number of basis functions $N$ and the number of bits of precision $m$ for representing the DVR matrix elements.  White equal volume isoline is shown. Blue color denotes region of advantage of the recursive method.}
    \label{fig:volume}
\end{figure}

The recursive method involves approximately $\mathcal{O}(4Fm^2+\frac{N^2}{F})$ T-gates compared to $\mathcal O(N\sqrt m)$. Choosing $F=\frac{N}{m}$ leads to $\mathcal{O}(Nm)$ T-gate cost for the recursive method, which is typically greater cost than that of the SELSWAP method. However the recurisive method is more favourable in terms of qubit count: $2n+9m$ compared do $N\sqrt{m}$. The quantum volume for the recursive method with $F=\frac{N}{m}$ is $\mathcal{O}(Nm^2)$ compared to $\mathcal{O}(N^2m)$ for the LKS method. 
The advantage of the recursive method is that it uses $\mathcal{O}(m)$ ancilla qubits and $\mathcal{O}(Nm^2)$ Clifford gates, compared to  $\mathcal{O}(N\sqrt{m})$ ancilla qubits and $\mathcal{O}(N^2m)$ Clifford gates for the LKS method. For the recursive-LKS method the quantum volume is reduced with respect to the SELECT-based recursive method by a constant factor. Quantum volume comparison between the SELSWAP (LKS) method and the recursive method is shown in Figure~\ref{fig:volume}. It follows that for DVRs of size greater than $2^7$ and at appropriately low precisions $m$ for representing the matrix elements, the recursive method is advantageous over the LKS method with respect to the quantum volume metric. 

We note that error-propagation when using arithmetics requires a qubit overhead of at least $2F$ additional qubits for representing $x$, $T_{pq}$ (for precision of $m$-bits for the final results).  It is thus a matter of choice and a specific use-case which method to choose. For optimizing the number of qubits used it is prerefable to use the recursive method, whereas for high-precision and low matrix size computations we recommend using the LKS method. 

\section{The DVR unitary}
\label{sec:DVRunitary}
\subsection{Construction via reflections}
\label{sec:DVRunitary-reflection}
The DVR unitary can be constructed from the DVR oracle using a technique given in ref.~\cite{Low2024}. The resulting unitary circuit represents the following transformation:
\begin{equation}
    \hat{D}=\ket{0}\bra{1}\hat{T}+\ket{1}\bra{0}\hat{T}^\dagger
    \label{eq:DVRunitary-reflection}
\end{equation}
The construction assumes providing an oracle returning the $k$-th column of the DVR unitary, $\ket{u_k}= \hat{T}\ket{k}$. Let $U_k$ prepare the following state:
\begin{equation}
	\ket{w_k}=\ket{0}_a\ket{k}-\ket{1}_a\ket{u_k}
	\label{eq:reflections_state}
\end{equation}
where $\ket{k}$ represents index-state labeling columns of the DVR matrix. Then the product of $N$ reflections:
\begin{equation}
    \hat{R}_k=I-2\ket{w_k}\bra{w_k}
    \label{eq:reflections}
\end{equation}
gives $\hat{D}=\Pi_{k=1}^N \hat{R}_k = \ket{0}\bra{1}\hat{T}+\ket{1}\bra{0}\hat{T}^\dagger$. Having execution of $\hat{T}$ and $\hat{T}^{\dag}$ conditioned on the state of ancilla qubit can be useful in certain scenarios, as discussed further. 
The total cost for the $\hat{D}$ unitary is $2N^2+N(4m+1)n$, where $m$ is the precision of matrix elements. This cost consists of the state preparation cost and the reflections cost. In the following paragraphs we give a more detailed resource estimation for the procedure sketched above for constructing $\hat{D}$.

\paragraph{State preparation.}

Let us consider the cost of preparing the state $\ket{w_k}$ given in eq.~\ref{eq:reflections_state}. For simplicity, consider the case of DVRs generated by polynomials with definite parity, i.e. $A_q=0$ in eq.~\ref{eq:dvr-recurrence}. For general polynomials the gate cost is roughly doubled due to twice the number of angles that must be loaded. Our present state preparation procedure follows the approach of Ref.~\cite{Low2024} and references therein.

{\bf Step 1. } First, initialize state $\ket{0}_a\ket{0}$, on which execute $HA_k \otimes I $ and get $\ket{0}_a\ket{0}\pm\ket{1}_a\ket{0}$, where $A_k\in\{I,X\}$ (if the first number in the $k$-th column of the $\mathbf{T}$ matrix is nonnegative we choose $A_k=X$, otherwise $A_k=I$. Next, create the state $\ket{0}_a\ket{k}\pm\ket{1}_a\ket{0}$ with controlled-$\hat{X}$ gates, with controls on the ancilla register $\ket{0}_a$ and $\log(N)$ target qubits for representing the integer index $\ket{k}$.

{\bf Step 2 } Consider the $n$-qubit state $\ket{.}=\ket{.}_0\otimes\ldots \otimes\ket{.}_{n-1}$ initialized for representing $\ket{u_k}$ given in eq.~\ref{eq:reflections_state}. We act on the $\ket{.}_0$ qubit with controlled-$H$ gate, where the condition is on the ancilla qubit  $\ket{.}_a$. Controlled Hadamard is equivalent to two T-gates. 

{\bf Step 3} Apply $CX^{n-1}$ on qubits $\ket{.}_1$,$ \ldots $,$\ket{.}_{t-1}$, with control on the $0$-th qubit to be $\ket{1}_0$. This is required for accounting for the horizontal symmetry of the DVR matrix. We reflect the elements in the upper half of the DVR matrix.

{\bf Step 4 } For a given $k$ denoting column index, follow the recursive procedure: for $t=1,\ldots,n-1$ load suitable angles $\varphi_{l}$ represented with $m+1$-bits: 
    \begin{equation}
        \hat{D}^{(t)}_{N,m+1}(\varphi)\ket{l}\ket{0}=\ket{l}\ket{\varphi_l}
        \label{eq:recursive-qrom}
    \end{equation}
where  $l=l_1l_2\ldots l_{t-1}$ is index

i.e. we load $2^{t-1}$ $(m+1)$-bit numbers. The extra $m+1$'th bit stores information about the sign of the coefficient. The QROM given in eq.~\ref{eq:recursive-qrom} is executed conditionally on the ancilla qubit $\ket{0}_a$ in state $\ket{1}_a$. Also, accounting for the horizontal symmetry in the DVR matrix we can ignore qubit $\ket{.}_0$ - the appropriate reflection is implemented in step 3. The angles $\varphi_l$ are chosen such that
\begin{equation}
   \cos^2 \varphi_l=\frac{\sum_{s=0}^{2^{-t-1}N-1}|T_{l2^{-t}N+s\; k}|^2}{\sum_{s=0}^{2^{-t}N}|T_{l2^{-t}N+s\; k}|^2} 
\end{equation}
i.e. they reproduce the correct probabilities in the $t$'th recursive step:
\begin{equation}
V_k^{(t)}\ket{0}_t=\cos|\varphi_l|\ket{0}_t+\sin|\varphi_l|\ket{1}_t
\end{equation}
and the relative phase between the elements is retrieved by loading one bit informing about the sign of  $\varphi_l$ and executing controlled-Z operation on qubit $\ket{\cdot}_t$. We choose  $\varphi_l$ to be nonnegative for $T_{l2^{-t}Nk}$, $T_{l2^{-t}N+2^{-t-1}Nk}$ both nonnegative or both nonpositive, and negative otherwise.

{\bf step 5.} Uncompute the $CX^{n-1}$ operation applied in step 3, If $k$ is odd, apply $Z$ to qubit $\ket{}_0$.

The Toffoli cost of the $t$-th recursive step in the procedure outlined above is $2^{t-1}$, giving the total cost
\begin{equation}
    \sum_{t=1}^{n-1} 2^{t-1} = 2^{n-1}-1 = \frac{1}{2}N-1
\end{equation}
in each recursive step the controlled single-qubit arbitrary rotation consumes $2m$ Toffoli gates (using phase-gradient technique). The circuit must be uncomputed adding another $\sum_{t=1}^{n-1} 2^{t-1} =  \frac{1}{2}N$ Toffoli gates, to give the final cost
 of the state preparation
 \begin{equation}
     C(\hat{U}_k) = N+2(n-1)m
 \end{equation}

\paragraph{Reflections.}

Implementing reflections $R_k=I-2\ket{w_k}\bra{w_k}$ requires loading $N/2$ angles that encode appropriate column of the DVR matrix $\ket{u_k}$. When $u_k$ is non-symmetric with respect to the midpoint (for orthogonal polynomials of indefinite parity), $N$ angles must be loaded. For the $\ket{0}\ket{k}$ part of the reflection no additional Toffoli gates are required, leaving $N/2+1$ Toffoli gates for state preparation with the SELECT-QROM method, plus equal number of Toffoli gates for uncomputation. This final Toffoli cost for the state preparation stage is thus $N+2m n$, where $2mn$ is from controlled-adder in the phase-gradient technique~\cite{Low2024}. Since $\hat{R}_k=U_k(I-2\ket{0}\bra{0})U_k^\dagger$ the cost implementing a single reflection operator is $2N+(4m +1)n$. Finally, the Toffoli gate cost of the DVR unitary $\hat{D}$ is

\begin{equation}
C_{REFL.}(\hat{D}) = \mathcal{O}\left(2N^2+N(4m+1)n\right).
    \label{eq:cost-D}
\end{equation}

\subsection{Construction via block-encoding and QROM}
\label{sec:DVRunitary-be}
The DVR unitary can be implemented via an construction alternative to the reflections method discussed in sec.~\ref{sec:DVRunitary-reflection}. The other technique relies on block-encoding of an appropriate matrix loaded with the following oracle: 
\begin{equation}
	\tilde{T}\ket{p}\ket{q}\ket{0}=\ket{p}\ket{q}\ket{\frac{2}{\pi}\arcsin T_{pq}}
	\label{eq:arcsin_oracle}
\end{equation}
The DVR unitary can be synthesized using the technique from ref.~\cite{camps}, giving $\frac{1}{N}{\hat{T}}$, followed by amplitude amplification~\cite{Brassard2002}, associated with an $N$-fold increase in the Toffoli cost. The SELECT circuit for $\tilde{T}$ costs $N$ Toffoli gates ($N^2/2$ for symmetric $T_{pq}=(-1)^qT_{N-pq}$ matrix elements). To get $\frac{1}{N^2}{\hat{T}}$ we construct the circuit implementing the following controlled-rotation:

\begin{equation}
P\ket{\frac{2}{\pi}\arcsin a} \ket0=\ket {\frac{2}{\pi}\arcsin a}(a\ket 0 +\sqrt{1-a^2}\ket1)
    \label{eq:controlled-rot-be}
\end{equation}
that involves $m$ controlled single-qubit rotations and costs roughly $m+m^2$  $(m-1)/2$ Toffoli gates. The gate count can be reduced at the expense of an increased number of qubits. Thus the Toffoli cost for implementing the DVR unitary scales as
\begin{equation}
C_{BE}(\hat{D}) = \mathcal{O}\left(N(N^2 +m^3)\right).
    \label{eq:cost-D-arc}
\end{equation}

\subsection{Construction via block-encoding and quantum arithmetics}
\label{sec:DVRunitary-be-art}
An alternative construction of $\tilde{T}$ uses quantum arithmetic operations to build matrix elements $\frac{2}{\pi}\arcsin T_{pq}$ defined in Eq.~\ref{eq:arcsin_oracle}. The benefit relative to the direct loading of the matrix elements with QROM depends on the magnitude of the entries in $\tilde{T}$ and their precision. For example, for the Gauss-Hermite DVR the largest $T_{pq}$ entry is around $0.5$, such that the Taylor series approximation $\arcsin(x) \approx x + x^{3}/6 + 3x^5/40 + 5x^7/102 + \frac{7!!}{8!!\cdot 9}x^9$ is sufficient for a 16-bit precision representation, with the associated cost of roughly $16m^2$ Toffoli gates.

In general, the construction of states storing values of a polynomial $P(x) = a_1x + a_3x^3 + \ldots + a_{2p-1}x^{2p-1}$ with quantum arithmetic operations requires $3pm^2 + \mathcal{O}(pm)$ Toffoli gates and $\mathcal{O}(mp)$ qubits. The procedure begins by initializing qubit registers in the states: $\ket{x}$, $\ket{x^2}$, $\ket{x^3}$, $\ket{x^5}$, ..., $\ket{x^{2p-1}}$, with the associated cost of $2pm^2$ Toffoli gates. Such a construction involves transforming the register $\ket{x}$ while keeping previously loaded numbers, by multiplying by $x^2$ to get $\ket{x^3}$ and so on. Each multiplication is associated with an approximate cost of $m^2$ Toffoli gates.

Next, using a separate qubit register, create $\ket{a_{2k-1}x^{2k-1}}$ for $k=1,2,...p$ and sum the terms, giving an extra cost of $m^2 + m$ Toffoli gates, per term. Thus, overall, including the cost of QROM for loading the initial matrix elements $T_{pq}$, the total cost of the quantum arithmetic method is 
\begin{equation} 
C_{art} = N^2 + 3pm^2 + \mathcal{O}(pm)
\end{equation} 
where $p$ depends on the desired accuracy and the type of Gaussian DVR. In practice, it can be net-beneficial to set a threshold value $\tau$ for the matrix elements $T_{pq}$, below which quantum arithmetic is used in constructing $\tilde{T}$, and QROM is used otherwise.

\section{Summary}
We developed a quantum algorithm that implements the discrete-variable representation (DVR) transformation. Our approach involves constructing a DVR oracle, which is then used to synthesize a unitary circuit implementing the DVR gate. We explore several synthesis strategies for the DVR gate, including a method based on reflections, block-encoding using QROM, and block-encoding with quantum arithmetic.

For the DVR oracle construction, we found that for matrix sizes of $N=2^7$ and larger, our recursive procedure (described in Section~\ref{sec:DVRoracle-rec}) can be advantageous to QROM, particularly when high precision is not required (see Fig.~\ref{fig:volume} and Section~\ref{sec:DVRoracle-compare}). For instance, for a 10-qubit DVR matrix at 16 bits of precision, the recursive procedure yields a gate count roughly 50\% lower than the QROM-based construction. However, it should be noted that this estimate does not account for numerical error propagation. Incorporating numerical error-propagation in the recursive construction of the DVR gate incurs an overhead of at least $2F$ qubits, which may offset the benefits over QROM. $F$ is the segmentation parameter defined in section~\ref{sec:DVRoracle-rec}.

We also presented two methods for implementing the DVR unitary gate. The first leverages a modified DVR oracle $\tilde{T}$ defined in eq.\ref{eq:arcsin_oracle}, followed by appropriate quantum rotations. The second, based on reflections, is discussed in Section~\ref{sec:DVRunitary-reflection} and implements Eq.~\ref{eq:DVRunitary-reflection}. The reflection-based method exhibits more favorable Toffoli gate count scaling with respect to matrix size and precision compared to the method using $\tilde{T}$: $\mathcal{O}\left(2N^2+N(4m+1)n\right)$ vs $\mathcal{O}\left(N(N^2 +m^3)\right)$, respectively.

In summary, we examined multiple strategies for constructing both the DVR oracle and the unitary DVR-FBR transformation circuit. Among these, we proposed a construction exploiting the recursive properties of orthogonal polynomials underlying the Gaussian DVR, which can outperform SELECT-SWAP QROM for large matrices and low precision. For synthesizing the DVR unitary circuit, we recommend the reflection-based technique described in Section~\ref{sec:DVRunitary-reflection}.

The advantages of implementing the DVR transformation directly within a quantum circuit are apparent in applications such as block-encoding many-body Hamiltonians, relevant to simulations of vibrational and vibronic dynamics in molecules and materials. In such Hamiltonians, the kinetic energy operator is naturally represented in a variational (FBR) basis, while the potential energy function is best described in a DVR basis, in which it is diagonal. Performing the FBR-to-DVR transformation on a $D$-particle bosonic product basis incurs a cost that scales linearly with $D$, relative to the 1D DVR transformation discussed here.
In contrast to grid-based and finite-difference methods, Gaussian DVRs exhibit exponential convergence of the approximation error with respect to basis set size. This arises from the exactness of the underlying $N$-point quadrature in integrating polynomials up to degree $2N-1$. As a result, DVR methods are highly effective for representing smooth wavefunctions of bound states that are common in many-body systems such as molecular vibrations.

This favorable scaling in both solution accuracy and quantum gate complexity (due to the tensor-product structure of multidimensional DVRs and the typically sparse nature of DVR Hamiltonians) makes the approach particularly appealing for quantum computation.

\section{Acknowledgments}
We thank Witold Jarnicki and Konrad Deka for helpful discussions.
This work is funded by the European Innovation Council accelerator grant COMFTQUA, no. 190183782. 

\section{Appendix A: Gaussian DVRs}
\label{sec:AppendixA}
In this appendix we discuss some of the useful properties of Gaussian DVRs~\cite{Stoer1980}, including the derivation of the diaognal form of local operators in DVR, a method of constructing the DVR matrix and the Schroedinger equation in multi-dimensional DVR~\cite{Light2000}. 

\paragraph{The potential energy operator in DVR.}
The DVR basis functions generally take non-zero values between quadrature points, however they tend to be smaller with increasing distance from the point of localization of the DVR function, as shown in Figure \ref{Fig:DVRbasis}. This observation intuitively supports the diagonal approximation to the potential matrix $V(x)$:
  \begin{equation}
  \begin{split}\label{eq:dvrkey}
V^{DVR}_{ij}=\langle d_i(x)|V(x)|d_j(x)\rangle=\int_a^bd_i(x)V(x)d_j(x)dx\approx\\
\approx \sum_{k=0}^{N-1}\frac{w_k}{\omega(x_k)}d_i(x_k)V(x_k)d_j(x_k)
=\sum_{k=0}^{N-1}\frac{w_k}{\omega(x_k)}\sqrt{\frac{\omega(x_k)}{w_i}}\delta_{ik}V(x_k)\sqrt{\frac{\omega(x_k)}{w_j}}\delta_{jk}=V(x_i)\delta_{ij}
\end{split}
\end{equation}
For the potential energy operators represented as
polynomial functions of coordinates (quadratic or higher degrees), the Gaussian quadrature
rule is not exact and the FBR/DVR approximation can no longer be considered variational.

\paragraph{DVR defined via the Position operator.}
A practical way of obtaining the DVR can be achieved through the so called \textit{product approximation} \cite{Light2000}, by means of diagonalisation of the position operator $X$ matrix. 
In VBR the matrix elements of $X$ are written as:
\begin{equation}\label{X}
 \left(\mathbf{X^{VBR}}\right)_{ij}=\int_a^b\phi_i(x)x\phi_j(x)dx\approx \sum_{k=0}^{N-1}\frac{w_k}{\omega(x_k)}\phi_i(x_k)x_k\phi_j(x_k)=\sum_{k=0}^{N-1}\mathbf{T}_{ik}^{T}\mathbf{X}_k\mathbf{T}_{kj}
\end{equation}
As long the $ \left(\mathbf{X^{VBR}}\right)_{ij}$ matrix is truncated to size $N \times N$ and the basis functions are orthogonal polynomials of degree $N-1$, the integrated function is of degree $2N-1$, and can be evaluated exactly by a Gaussian quadrature. It means that the $\mathbf{T}$ matrix diagonalizes the position operator matrix in the orthogonal polynomials basis. As a result,  eigenvalues of $\mathbf{X}$ in this basis correspond to quadrature points and the diagonalising  transformation matrix is related to the quadrature weights. Diagonalisation of $\mathbf{X^{VBR}}$ unambiguously defines the DVR:
\begin{equation}\label{X1}
 \mathbf{X^{DVR}}=\mathbf{T}\mathbf{X^{VBR}}\mathbf{T}^{T}
\end{equation}
where we used the fact that for the Gaussian quadratures the FBR-DVR transformation matrix is unitary: $\mathbf{T}^{-1}=\mathbf{T}^{T}$.
The VBR representation of the position matrix in the orthogonal polynomials basis is straightforward to derive, thanks to the three-term recurrence relations for orthogonal polynomials: $p_{q+1}(x)=(A_q+B_qx)p_q(x)+C_qp_{q-1}(x)$, where $A_q,B_q,C_q$ are constants characteristic for a given class of polynomials. As a result the position operator matrix is tridiagonal. In practice, diagonalising the position operator matrix is the most efficient way of finding a DVR.

\begin{figure}
\begin{center}
  \includegraphics[width=15cm]{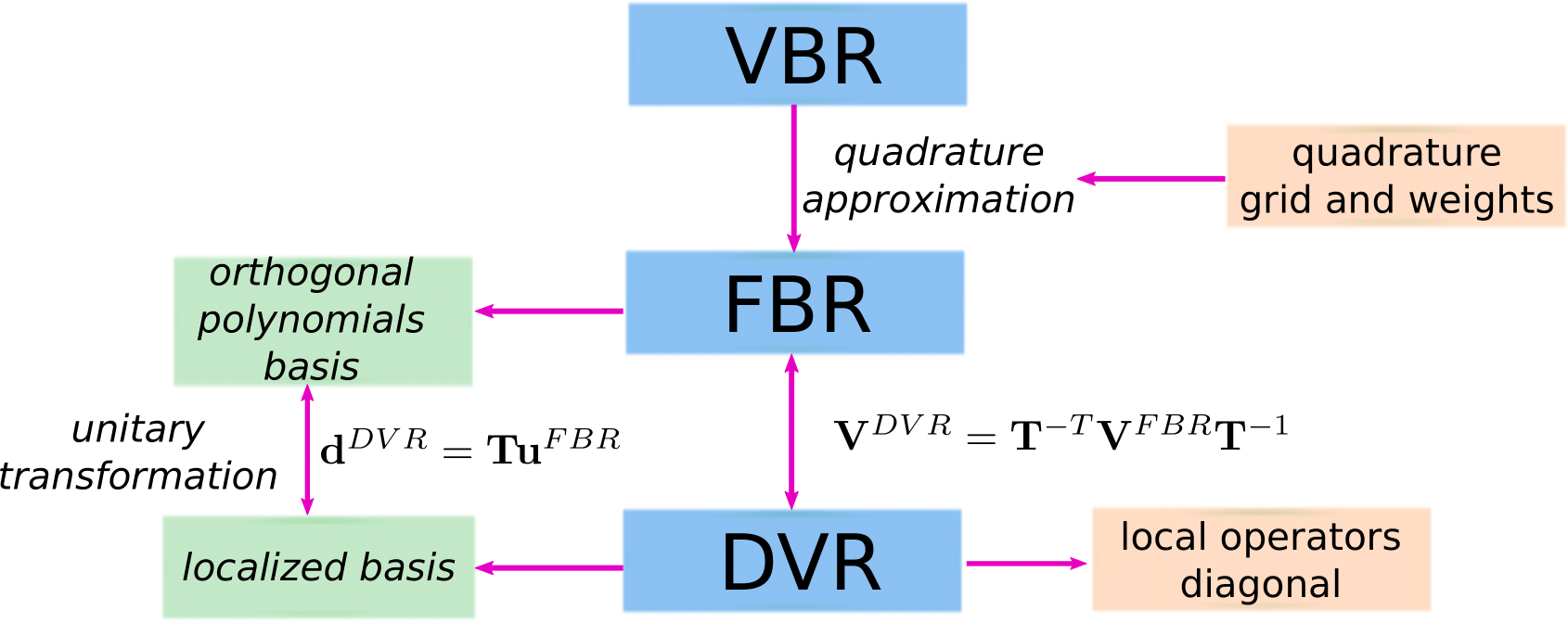}\\
  \caption{A general scheme for the VBR-FBR-DVR transformation. $\mathbf{V}^{FBR}$ denotes FBR representation of a local operator and $\mathbf{T}$ is the FBR-to-DVR transformation matrix.}\label{fig:VBR-FBR-DVR}
  \end{center}
\end{figure}

\paragraph{The Schroedinger equation in DVR.}
The Hamiltonian given in the variational basis (VBR) can be approximated with finite basis representation as follows
\begin{equation}
\mathbf{H}^{VBR}\approx \mathbf{H}^{FBR}=\mathbf{K}^{VBR}+\mathbf{T}^{T}\mathbf{V}^{diag}\mathbf{T}
\label{eq:FBR1D}
\end{equation}
 where the kinetic energy operator $\mathbf{K}^{VBR}$ remains in the VBR representation and quadrature approximation is applied to the potential energy matrix elements $\mathbf{V}^{VBR} \approx \mathbf{T}^{T}\mathbf{V}^{diag}\mathbf{T}$. Then the Schroedinger equation can be written in the matrix form as:
 \begin{equation}
\left(\mathbf{K}^{VBR}+\mathbf{T}^{T}\mathbf{V}^{diag}\mathbf{T}\right)\mathbf{U}=\mathbf{T}^T\mathbf{T}\mathbf{U}\mathbf{E}
\label{eq:FBR1DSE}
\end{equation}
where overlap integral matrix $S$ on the right-hand side is also given in quadrature approximation, i.e. $\mathbf{S}=\mathbf{T}^T\mathbf{T}$.
Upon multiplication of eq.~\ref{eq:FBR1DSE} from the left by $\mathbf{T}^{-T}$ and defining a new basis by $\mathbf{Z}=\mathbf{T}\mathbf{U}$ we get the DVR form of the Schroedinger equation:
\begin{equation}
\left(\mathbf{T}^{-T}\mathbf{K}^{VBR}\mathbf{T}^{-1}+\mathbf{V}^{diag}\right)\mathbf{Z}=\mathbf{Z}\mathbf{E}
\label{eq:DVR1DSE}
\end{equation}
A schematic for the VBR-FBR-DVR transformation chain is depicted in Figure~\ref{fig:VBR-FBR-DVR}.

This scheme extends to many-dimensions by incorporating direct-product basis set representatino of the multi-dimensional problem. The D-dimensional basis functions gain $D$ indices as follows: $\lbrace \phi_{n_1,n_2,...,n_D} := \phi_{n_1}(Q_1)\phi_{n_2}(Q_2)...\phi_{n_D}(Q_D)\rbrace_{n_c=1,2,...,N_c; c=1,..,D}$. Direct product quadrature grid is a grid composed as a simple sum of 1-dimensional grids so that the indices of the multidimensional grid are independent. The multi-dimensional FBR Schroedinger equation then takes the form
\begin{equation}
\left(\mathbf{K}^{VBR}+\mathbf{T}^{T}\mathbf{V}^{diag}\mathbf{T}\right)\mathbf{U}=\mathbf{T}^T\mathbf{T}\mathbf{U}\mathbf{E}
\label{eq:FBRDDSE}
\end{equation}
where $\mathbf{T}=\mathbf{^{(1)}T}\otimes\mathbf{^{(2)}T}\otimes...\otimes\mathbf{^{(D)}T}$ with $\mathbf{^{(c)}T}_{p_cq_c}=N_{q_c}\sqrt{\mathbf{^{(c)}w}_{k_cn_c}}\mathbf{^{(c)}p}_{p_cq_c}$, where $\mathbf{^{(c)}p}_{p_cq_c}=p_{q_c}(Q_{p_c})$.  The DVR representation is then obtained by transposing the Kronecker $\mathbf{T}$ directly, at no additional cost compared to the one-dimensional problem. 
Note that the size of the wavefunction $\mathbf{U}$ is $N^D$ which quickly becomes a prohibitively large number with increasing $D$. For instance, if 10 basis functions per dimension are needed  to converge the variational wavefunction, then for a 5-atom molecule ($D=9$) e.g. CH$_4$, as many as $10^9$ basis functions are required. For molecules with more than 5 atoms, the amount of memory needed to store even a single wavefunction vector becomes problematic. To circumvent this \textit{curse of dimensionality} several approaches have been developed to define FBR and DVR with non-direct product basis sets and non-direct product quadrature grids~\cite{gaborig,AvCaV,Zak2019,Avila2017ch2nh}. With quantum computer qubit encoding of information, this memory requirement can be alleviated to $\mathcal{O}(ND)$ qubits, while the tensor structture of the multidimensional DVR transformation allows to execute this transformation at low depth (each dimension can be transformed simultaneously).

\section{Appendix B: orthogonal polynomials}
\label{sec:appendixB}
In this appendix we lay out properties of popular classes of orthogonal polynomials used in construction of Gaussian DVRs.

{\bf I. Generalized Laguerre polynomials}
For fixed $\alpha\in\mathbb{R}$ orthogonal polynomials with respect to $\mu=x^\alpha e^{-x}\chi_{[0,\infty)}$ 

We have the formula$$L_n^{(\alpha)}(x)=\frac{x^{-\alpha}e^{x}}{n!}(x^{n+\alpha}e^{-x})^{(n)}=(-1)^nx^n/n!+\ldots \,,$$
the recurrence relations
$$L_n^{(\alpha)}(x)=\frac{(-x+2n+\alpha-1)L_{n-1}^{(\alpha)}(x)-(n+\alpha-1)L_{n-2}^{(\alpha)}(x)}{n}$$
and the norm $$\|L_{n}^{(\alpha)}\|_{L^2(\mu)}=\sqrt{\frac{\Gamma(\alpha+n+1)}{\Gamma(n+1)}}\,.$$
Thus $$t_{pq}=\frac{(\alpha+q)^{1/2}(-x+2q+\alpha-1)t_{pq-1}}{q^{3/2}}-\left(\frac{(q+\alpha-1)}{q}\right)^{3/2}\sqrt{\frac{q+\alpha}{q-1}}t_{pq-2}\,,$$

{\bf  Laguerre polynomials: $\alpha=0$}
Orthogonal polynomials with respect to $\mu= e^{-x}\chi_{[0,\infty)}$ 

We have the formula$$L_n(x)=L_n^{(0)}(x)=\frac{e^{x}}{n!}(x^{n}e^{-x})^{(n)}=(-1)^nx^n/n!+\ldots \,,$$
the recurrence relations
$$L_n(x)=\frac{(-x+2n-1)L_{n-1}(x)-(n-1)L_{n-2}}{n}$$
and the norm $$\|L_{n}\|_{L^2(\mu)}=1\,.$$
Thus $$t_{pq}=\frac{(-x_p+2q-1)}{q}t_{pq-1}-\frac{(q-1)}{q}t_{pq-2}\,,$$

{\bf II. Jacobi polynomials}
For fixed $\alpha,\beta>-1$ orthogonal polynomials with respect to $\mu=(1-x)^\alpha(1+x)^\beta\chi_{[-1,1]}$ .

We have the formula
$$P_n^{(\alpha,\beta)}(x)=\frac{(-1)^n}{2^nn!}(1-x)^{-\alpha}(1+x)^{-\beta}\left((1-x)^\alpha(1+x)^\beta(1-x^2)^n\right)^{(n)}$$
$$=\sum_{s=0}^n\binom{n+\alpha}{n-s}\binom{n+\beta}{s}\left(\frac{x-1}{2}\right)^s\left(\frac{x+1}{2}\right) ^{n-s}\,,$$
the recurrence relations
$$2n(n+\alpha+\beta)(2n+\alpha+\beta-2)P_n^{(\alpha,\beta)}(x)$$
$$=(2n+\alpha+\beta-1)\left((2n+\alpha+\beta)(2n+\alpha+\beta-2)x+\alpha^2-\beta^2\right)P_{n-1}^{(\alpha,\beta)}(x)$$
$$-2(n+\alpha-1)(n+\beta-1)(2n+\alpha+\beta)P_{n-2}^{(\alpha,\beta)}(x)$$
and the norm $$\|P_n^{(\alpha,\beta)}(x)\|_{L^2(\mu)}=\sqrt{\frac{2^{\alpha+\beta+1}}{2n+\alpha+\beta+1}\frac{\Gamma(\alpha+n+1)\Gamma(\beta+n+1)}{\Gamma(\alpha+\beta+n+1)n{!}}}\,.$$
Thus recurrence relations for $t_{pq}$ are complicated (but we can write it without $\Gamma$ function).

It seems hard to deal with it but let us look at some special cases.

{\bf  Legendre polynomials: $\alpha=\beta=0$}
Orthogonal polynomials with respect to $\mu= \chi_{[-1,1]}$ 

We have the formula
$$P_n=P_n^{(0,0)}(x)=\frac{1}{2^nn!}\left((x^2-1)^n\right)^{(n)}$$
the recurrence relations
$$P_n(x)=(2-1/n)xP_{n-1}(x)-(1-1/n)P_{n-2}$$
and the norm $$\|P_{n}\|_{L^2(\mu)}=\sqrt{\frac{2}{2n+1}}\,.$$
Thus $$t_{pq}=\frac{\sqrt{4q^2-1}}{q}x_pt_{pq-1}-\frac{(q-1)}{q}\sqrt{\frac{2q+1}{2q-3}}t_{pq-2}\,,$$

{\bf  Chebyshev polynomials (second kind): $\alpha=\beta=1/2$}
Orthogonal polynomials with respect to $\mu= \sqrt{1-x^2}\chi_{[-1,1]}$ 

We have the formula
$$U_n(x)=P_n^{(1/2,1/2)}(x)=\sum_{k=0}^{[n/2]}(-1)^k\binom{n-k}{k}(2x)^{n-2k},\;\;U_n(\cos t)=\frac{\sin\left((n+1)t\right)}{\sin t}\,,$$
the recurrence relations
$$U_n(x)=2xU_{n-1}(x)-U_{n-2}$$
and the norm $$\|U_{n}\|_{L^2(\mu)}=\sqrt{\pi/2}\,.$$
Thus $$t_{pq}=2x_pt_{pq-1}-t_{pq-2}\,,$$
$$\gamma_q=(-1)^{[(q+1)/2]}$$
and
$$t_{pq}'=(-1)^q2x_pt_{pq-1}'+t_{pq-2}'$$ thus we do not need to multiply by $\gamma_q$.

{\bf  Chebyshev polynomials (first kind): $\alpha=\beta=-1/2$}
Orthogonal polynomials with respect to $\mu= (1-x^2)^{-1/2}\chi_{[-1,1]}$ 

We have the formula
$$T_n(x)=P_n^{(-1/2,-1/2)}(x)=\sum_{k=0}^{[n/2]}\binom{n}{2k}(x^2-1)^kx^{n-2k},\;\;T_n(\cos t)=\cos\left(nt\right)\,,$$
the recurrence relations
$$T_0=1,\;T_1(x)=x,\;T_n(x)=2xT_{n-1}(x)-T_{n-2}$$
and the norm $$\|T_{0}\|_{L^2(\mu)}=\sqrt{\pi} \hbox{ and }\|T_{n}\|_{L^2(\mu)}=\sqrt{\pi/2} \hbox{ for } n>0\,.$$
Thus $$t_{p2}=2x_pt_{p1}-\sqrt{1/2}t_{p0} \hbox{ and } t_{pq}=2x_pt_{pq-1}-t_{pq-2} \hbox{ for } q>2\,,$$
$$\gamma_q=(-1)^{[(q+1)/2]}\sqrt{(3+(-1)^q)/2}$$
and
$$t_{pq}'=(-1)^q2x_pt_{pq-1}'+t_{pq-2}'$$ thus again we do not need to multiply by $\gamma_q$.

\section{Appendix C: Quantum arithmetic for Quantum DVR (QDVR)}
\label{sec:appendixC}
In this appendix, we discuss details of quantum arithmetic required for the implementation of the unitary blocks $U_c$ in the recursive scheme for constructing the DVR oracle shown in Figure~\ref{fig:DVRoracle}. The dominant cost arises from the multiplication of numbers, which requires $\mathcal{O}(m^2)$ gates in each $U_c$.

Recalling eq.~\ref{eq:uc-unitaries2}:
\begin{align}
U_{2c} \ket q\ket x\ket{A}\ket{B} &=
    \ket q\ket x\ket{A} \ket{B\oplus (A'_{\hat q}+B'_{\hat q}x)A}  
    \label{eq:uc-unitaries1-app}
    \end{align}
we need to multiply the numbers $x$, $A$, and $B'_q$. Multiplication by $B'_q$ is cheaper than by $A$ because $B'_q$ is a fixed number (not a quantum input). These numbers are multiplied as integers written in binary system. The cost of computing $xA$ is approximately $2\,{\rm length}(x)\,{\rm length}(A)$ Toffoli gates, plus additional Clifford gates. Here ${\rm length}(x)$ corresponds to binary precision, commonly denoted with $m$ throughout the text. The cost of multiplying $xA$ by $B'_q$ is ${\rm length}(xA)\cdot{\rm length}(B'_q)$ Toffoli gates, plus Clifford gates. Of course, the exact cost depends on the multiplication algorithm; here, we assume the most basic textbook method, i.e. without using the Quantum Fourier Transform.

Next, we add $(xAB'_q)_m$, which denotes the $m$ most significant bits of $xAB'_q$, to $A'_q$, conditioned on the ancilla qubit that encodes $v' \geq c$ (see Figure~\ref{fig:arithmetic}). This controlled addition costs approximately $2m$ Toffoli gates, plus $\mathcal{O}(m)$ Clifford gates. Finally, we uncompute $xAB'_q$.

The basic algorithm we utilize for multiplying two numbers $a$ and $b$ is as follows. Let $a$ and $b$ be $m$-bit and $m'$-bit numbers, respectively, and let $m'' = m + m' + 1$. We perform $m$ controlled additions: the $i$-th Adder $A_i$ is controlled on the bit $a_i$ in the decomposition $a = \sum a_i 2^i$ ,for $i = 0, \ldots, m-1$:
$$
A_i\ket{a}\ket{b}\ket{c} = 
\begin{cases}
    \ket{a}\ket{b}\ket{c} & \text{if } a_i = 0 \\
    \ket{a}\ket{b}\ket{c + 2^i b} & \text{otherwise.}
\end{cases}
$$
Each such adder is an addition of two $m$-bit numbers: $b$ and $[c/2^i]$, and costs approximately $2m$ Toffoli gates. If we multiply by a fixed number, the operation costs (depending on the specific value) $2\times$ fewer Toffoli gates, since the Adders are then controlled by classical bits. We note that multiple possibilities for optimizing this cost may exist.

The multiplication operation $Mult_x\ket{A} = \ket{xA}$ can be implemented in the following way: if $x$ is an $m$-bit (nonzero) number and $a$ is $m'$-bit, then we can perform
$$
Mult_x\ket{A}_{m'}\ket{0}_{m'+1} = \ket{xA}_{m''}.
$$
which requires $m'$ additional ancilla qubits. Note that if $|x| \neq 1$, we cannot simply approximate $xA$ by truncating bits, since $Mult_x$ is a unitary operation and must therefore be injective. For example, if $x = 2^{-m}$, we require at least $2^{-(m + m')}$ precision. Consequently, if we perform $K$ such multiplications as in our construction, we need $\mathcal{O}(Nm)$ additional ancilla qubits.

\end{document}